\documentclass[11pt]{article}

\usepackage{epsf}
\usepackage{graphicx}
\usepackage{amssymb}
\usepackage{amsmath}

\font\cmr=cmr7

\newcommand{\be}{\begin{equation}}
\newcommand{\ee}{\end{equation}}
\newcommand{\bea}{\begin{eqnarray}}
\newcommand{\ena}{\end{eqnarray}}
\newcommand{\vs}[1]{\rule[- #1 mm]{0mm}{#1 mm}}

\newcommand{\pT}{{p_{_T} }}
\newcommand{\dsig}{\frac{d \sigma}{d {\vec p}_{_T} d \eta}}
\newcommand{\dsigd}{\frac{d \sigma^{\hbox{\cmr (D)}}}{{d {\vec p}_{_T} d \eta}}}
\newcommand{\dsigf}{\frac{d \sigma^{\hbox{\cmr (F)}}}{{d {\vec p}_{_T} d \eta}}}
\newcommand{\dsigij}{\frac{d {\widehat \sigma}_{ij}}{{d {\vec p}_{_T} d \eta}}}
\newcommand{\dsigijk}{\frac{d {\widehat \sigma}_{ij}^{k}}
                           {d {\vec p}_{_T} d \eta}}
\newcommand{\kd}{K^{\hbox{\cmr (D)}}}
\newcommand{\kf}{K^{\hbox{\cmr (F)}}}
\newcommand{\MSB}{{\overline {MS}}}
\newcommand{\LMS}{\Lambda_{_{_\MSB}}}

\newcommand{\alfspi}{\frac{\alpha_s(\mu_{R})}{2 \pi}}

\newcommand{\PR}[1]{Phys.\ Rev.\ {\bf #1}}

\begin{document}

\renewcommand{\thefootnote}{\fnsymbol{footnote}}
\newpage
\setcounter{page}{0}

\vs{10}
%\vs{40}
\date{today}
\begin{center}
%{\Large {\bf{A PHENOMENOLOGICAL APPRAISAL}}}\\
%or \\
{\Large {\bf{A NEW CRITICAL STUDY OF PHOTON}}} \\
\vspace{0.5 cm}
{\Large {\bf{PRODUCTION IN HADRONIC COLLISIONS}}} \\
%%%%%%\vspace{0.5 cm}
%%%%%%{\Large {\bf{}}} \\
\vspace{0.9 cm}
{\large P.~Aurenche$^1$, M.~Fontannaz$^2$, J.Ph.~Guillet$^1$, E.~Pilon$^1$, 
M.~Werlen$^1$} \\
\vspace{0.7 cm}
\end{center}
%\maketitle
\begin{itemize}
%\begin{enumerate}
\item[1.] {\em Laboratoire d'Annecy-le-Vieux de Physique Th\'eorique LAPTH,
\footnote{UMR5108 du CNRS associ\'ee \`a l'Universit\'e de Savoie.}\\
B.P. 110, F-74941 Annecy-le-Vieux Cedex, France} \\
%\vspace{0.7 cm}
\item[2.] {\em Laboratoire de Physique Th\'eorique LPT,
\footnote{UMR8627 du CNRS.}\\
B\^at. 210, Universit\'e de Paris-Sud, F-91405 Orsay Cedex, France} \\
\end{itemize}
%\end{center}

%\vs{7}
\vs{10}

\centerline{ \bf{Abstract}}
%\vs{5}
%\vs{3}
\noindent
In the light of the new prompt photon data collected by PHENIX at RHIC and by 
D$\emptyset$ at the run II of the Tevatron, we revisit the world prompt photon 
data, both inclusive and isolated, in hadronic collisions, and compare them 
with the NLO QCD calculations implemented in the Monte Carlo programme 
{\tt JETPHOX}.
%%%%% for one particle inclusive cross sections as well as correlations.
%%%%% Predictions are made for photon-jet correlations and also for hadron-jet
%%%%% correlations which should help better understand the parton fragmentation
%%%%% functions.

\vfill
\rightline{hep-ph/0602133}
\rightline{LAPTH-1140/06}
\rightline{LPT-Orsay/05-75}

\setcounter{footnote}{0}
\newpage
         
\section{Introduction}

Two experiments have recently collected prompt photon data. For the first time,
a collaboration (PHENIX at $\sqrt{s} = 200$~GeV at RHIC) has been able to
collect data both for the inclusive~\cite{Adler:2005} and the isolated
case~\cite{Okada:2005}  which should help better understand the role of the
fragmentation component in prompt photon production. Furthermore, during run II
of the Tevatron ($\sqrt{s} = 1.96$ GeV), the D$\emptyset$
collaboration~\cite{Abazov:2005wc} has measured isolated prompt photons whose
transverse momenta $p_{T}$ range from 23 to about 300 GeV, the widest domain
ever covered. These new experimental results shed some light on a controversy
which has been plaguing prompt photon phenomenology since the  late 90's.

\vspace{0.2cm}

\noindent
Indeed, many years of intense experimental efforts, ranging from fixed targets
\cite{EXP,R806,WA70,R110,R807,E706,E70604,UA6} to colliders
\cite{UA1,UA2,D02000,D02001,CDF2002,CDF2004}, have led to a wealth of experimental data on prompt
photon production in hadronic collisions, but also to a controversial
situation. In particular, in the late 90's some confusion was created by one
fixed target experiment \cite{E706} which found cross sections several times
above theoretical predictions based on Next-to-Leading Order (NLO)
calculations;
data and theory disagreed both in magnitude and shape. This disagreement
triggered a debate on large recoil effects possibly of non perturbative origin.
The resummation of threshold as well as recoil effects induced by soft gluon
radiation in the single particle inclusive cross sections have been performed.
In the meantime, full NLO calculations have been implemented in more flexible
Monte Carlo programmes at the partonic level. Programmes of this type account
for experimental cuts in an easy way, match naturally the binning of
experimental data and, by histogramming of the partonic
configurations generated, allow for a straightforward study of correlations.
The latter provide more constrained and refined tests of the short distance dynamics
than the single particle inclusive distributions.

\vspace{0.2cm}

\noindent
In the present article we propose a reexamination of prompt photon data in the 
light of these new experimental results. In section \ref{theory} we formulate 
the theoretical framework of our study, and discuss the complementary features
of dedicated resummed calculations and NLO calculations implemented in Monte
carlo programmes such as {\tt JETPHOX}~\cite{cocorico} used in the present study. In section
\ref{comp-data}, we present a comparison of this theoretical framework with the
new PHENIX data, and with the new D$\emptyset$ data. We then reexamine the older world
data in the light of this comparison. Section \ref{conclusions} contains our
conclusions.  

%%%  Section \ref{correl} is dedicated to a study of photon-jet and photon-hadron correlations.

%%%%%%%%%%%%%%%%%%%%%\section{Theoretical framework and ambiguities}\label{theory}
\section{Theoretical framework and ambiguities}\label{theory}

\subsection{Mechanisms of production of prompt photons}\label{incl} 

Schematically, the production of a prompt photon proceeds via two
mechanisms. In the first one, which may be called `direct' (D), the photon
behaves as a high $p_{T}$ colourless parton, i.e. it takes part in the hard
subprocess, and it is most likely to be well separated from any hadronic
environment. In the other one, which may be called `fragmentation' (F), the
photon behaves as a kind of (anomalous) hadron, i.e. it results from the
collinear fragmentation of a coloured high $p_{T}$ parton, and is it most
probably accompanied by hadrons - unless the photon carries away most of the
transverse momentum of the fragmenting parton, which is usually the situation
in fixed target experiments. 

\vspace{0.2cm}

\noindent
From a technical point of view, (F) emerges from the calculation of the higher
order corrections to (D) in the perturbative expansion in powers of the strong 
coupling $\alpha_{s}$. At higher orders, final state multiple collinear
singularities appear in any subprocess where a high $p_{T}$ parton of species 
$k$ (quark or gluon) undergoes a cascade of successive collinear splittings 
ending up with a splitting into a photon. These singularities are factorised to
all orders in $\alpha_s$ according to the factorisation theorem, and absorbed
into fragmentation functions of parton $k$ to a photon,  
$D_{\gamma/k}(z,M_{_F})$, defined in some arbitrary fragmentation scheme, at
some arbitrary fragmentation scale $M_{_F}$. The point-like coupling of the
photon to quarks is responsible for the well-known anomalous behaviour of
$D_{\gamma/k}(z, M_{_F})$, roughly as $\alpha_{em}.\alpha_{s}^{-1}(M_{_F})$
when the fragmentation scale $M_{_F}$, chosen of the order of a hard scale of
the subprocess, is large compared to ${\cal O}(1)$~GeV. In this article, (D) is
precisely given by the Born term  plus the fraction of the higher order
corrections from which final state collinear singularities have been subtracted
according to the ${\overline{\mbox{MS}}}$ factorization scheme. (F) is the
contribution involving a fragmentation function of any parton into a photon in
the ${\overline{\mbox{MS}}}$ factorization scheme. The differential cross
section in transverse momentum $\pT$ and rapidity $\eta$ can thus be written
synthetically as: 
\begin{equation} 
\sigma^{\gamma} 
=
\sigma^{(D)}(\mu_{R},M,M_{_F}) 
+\sum_{k=q,\bar{q},g}\sigma^{(F)}_{k}(\mu_{R},M,M_{_F}) 
\otimes D_{\gamma/k}(M_{_F}) 
\label{eq1}
\end{equation}
where $\sigma^{(F)}_{k}$ describes the production of a parton $k$ in a
hard collision. The arbitrary parameters $\mu_{R}$, $M$ and $M_{_F}$
are respectively the renormalisation, initial-state factorisation, and
fragmentation scales. Let us stress once more that the splitting between 
(D) and (F) is arbitrary: it relies on a choice of factorisation scheme and 
scale to which refer the definitions of each of these contributions. In
particular, both (D) and (F) depend on $M_{_F}$, so that the partial
cancellation of the $M_{_F}$ dependence in the predictions proceeds in a
qualitatively different, and quite more complicated, way here than in the 
purely hadronic case. The dependence of the NLO predictions with respect 
to $\mu_{R}$, $M$ and $M_{_F}$ will be discussed in sect. \ref{comp-data}.
When all scales are taken to be equal, they will be noted $\mu$.

\vspace{0.2cm}

\noindent
The study provided in this article relies on the calculation of both (D) and
(F) at next-to-leading order (NLO) accuracy \cite{cocorico}, which takes the
form ($\eta$ is the photon rapidity)
\be
\dsig = \dsigd\ + \ \dsigf
\label{eq:sig}
\ee
where 
\bea
\dsigd 
 = \sum_{i,j=q,\bar{q},g} \int dx_{1} dx_{2}
\ F_{i/h_1}(x_{1},M)\ F_{j/h_2}(x_{2},M) 
\alfspi \left( \dsigij + \alfspi \kd_{ij} (\mu_{R},M,M_{_F}) \right)
\label{eq:dir}
\ena
and
\bea
\dsigf &=& \sum_{i,j,k=q,\bar{q},g} \int dx_{1} dx_{2}\frac{dz}{z^2}
\ F_{i/h_1}(x_{1},M)\ F_{j/h_2}(x_{2},M)\ D_{\gamma/k}(z, M_{_F})
\nonumber \\
&\ & \qquad \qquad
\ \left( \alfspi \right)^2\  \left( \dsigijk \ + 
\ \alfspi \kf_{ij,k} (\mu_{R},M,M_{_F}) \right)
%%%%%%%%%%%%%%%%%%%%\dsigf = \sum_{i,j,k=q,\bar{q},g} \int dx_{1} dx_{2}\frac{dz}{z^2}
%%%%%%%%%%%%%%%%%%%%\ F_{i/h_1}(x_{1},M)\ F_{j/h_2}(x_{2},M)\ D_{\gamma/k}(z, M_{_F})
%%%%%%%%%%%%%%%%%%%%\ \left( \alfspi \right)^2\  \left( \dsigijk \ + 
%%%%%%%%%%%%%%%%%%%%\ \alfspi \kf_{ij,k} (\mu_{R},M,M_{_F}) \right)
\label{eq:brem}
\ena 
where $F_{i/h_{1,2}}(x,M)$ are the parton distribution functions of  parton
species $i$ inside the incoming hadrons $h_{1,2}$, at momentum fraction $x$ and
factorisation scale $M$; $\alpha_{s}(\mu_{R})$ is the strong coupling  defined
in the $\overline{\mbox{MS}}$ renormalisation scheme at the  renomalisation
scale $\mu_{R}$.  The knowledge of  $\LMS$, e.g. from deep-inelastic scattering
experiments, completely specifies  the NLO expression of the running coupling
$\alpha_{s}(\mu_{R})$. The NLO correction terms to (D) and (F),
$\kd_{ij}$~\cite{ABDFS,GV} and $\kf_{ij,k}$~\cite{ACGG} respectively, are known
and their expressions in the 
$\overline{\mbox{MS}}$ scheme will be used. The dependence of these functions
on the kinematical variables $x_1, x_2, z, \sqrt{s}, \pT$ and $\eta$ has
not been explicitly displayed. 
The structure and fragmentation functions have been determined 
at the required level of accuracy by NLO fits to the data.
%%%% All the quantities entering the above
%%%% equations have been either calculated ($\kd_{ij}$ and $\kf_{ij,k}$) or

\vspace{0.2cm}

\noindent
The results of the NLO calculation of (D) have been known for a long time 
\cite{ABDFS}. They were first implemented in computer codes in a form dedicated
to one particle inclusive distributions, the integration over  the  phase space
variables being done analytically. As such, these
codes were fast but not flexible enough to account for the various
experimental selection and, especially, the isolation cuts used at colliders,
and they were not suited to study correlations.
The calculation has been subsequently implemented using a `Monte Carlo' method
\cite{owens} which however included the (F) contribution only at leading order
(LO) accuracy. The calculation of the NLO corrections to (F) became also
progressively available along the same steps \cite{ACGG,gordon,cocorico}.
The present study relies on the implementation of the NLO calculation of both
(D) and (F) in a Monte Carlo  programme\footnote{For inclusive observables
involving no isolation cuts, we have also used the much faster NLO
programme {\tt INCNLO} \cite{cocorico} implementing analytic expressions of the
HO corrections.}, called  {\tt JETPHOX}, briefly described in
subsect.~\ref{jetphox}.

\vspace{0.2cm}

\noindent
More recently, expressions involving the resummation, at next-to-leading
logarithmic (NLL) accuracy of terms which are logarithmically large
at the phase space boundary ($x_{T} = 2 p_{T}/\sqrt{s} \to 1)$ have been
obtained, first in (D) \cite{los,cmn,cmnov,ko,sv}, and more recently in (F) as 
well \cite{deFlorian:2005wf}. This resummation is performed only for inclusive 
$p_{_T}$ distributions integrated over all rapidities. 
The effect of this resummation extends down to values of
$x_{T} \geq $ a few $10^{-1}$ and thus covers the range of fixed target
experiments. They provide a much reduced $\mu_{R}$ and $M$ dependence than
with the NLO approximation. The NLO results roughly agree with the resummed
calculation in the region populated by fixed target data, when $\mu_{R}$ and
$M$ are chosen  $\sim p_{T}/2$ in the former. 
%As discussed in sect.
%\ref{scale-dep}, such a  scale choice in the NLO approximation is
%independently motivated by a detailed  study of scale dependences. These are
%the scales at which the NLO result is the least sensitive to scale variations,
%providing the `optimal scale choice' in this respect \cite{optim}. 
As for
collider experiments, these resummations do not have much impact on the
phenomenology in the $x_{T}$ range  $\sim 10^{-2}$ to $10^{-1}$ covered by the
data. The impact of resumming logarithmically enhanced terms at small $x_{T}$ 
might be more relevant in the smaller $x_{T}$ range, yet a proper resummation of
this kind has not been studied so far in prompt photon production, to our 
knowledge.

\vspace{0.2cm}

\noindent
Even more recently, a joint summation of both threshold and recoil effects due
to soft multigluon emission has been performed in \cite{Sterman:2004yk}. Recoil effects 
are logarithmically enhanced order by order in the $\alpha_{s}$ expansion of
the $q_{T}$ distribution of a pair $\gamma$-jet; however this logarithmic
enhancement is washed out by the integration over the jet when passing to the
single photon inclusive $p_{T}$ distribution. The joint summation recently
performed confirmed that this order by order conclusion also holds in the all
order resummed result when resummation is performed {\it before} the
integration over the recoiling jet, leaving un-enhanced contributions only,
whose effects remain small. The joint summation makes also contact with
possible non perturbative effects of $k_{T}$ kick which are not accounted for 
in any fixed order calculation. However these non perturbative recoil effects, 
which are to a large extend unconstrained by theory so far, remain small 
unless large non perturbative parameters are used; one experiment only argues 
in favour of such unexpectedly large parameters. This issue will be discussed 
in sect.~\ref{comp-data}.        

\vspace{0.2cm}

\noindent
All these resummed calculations have to be performed in a space conjugate to
the physical phase space through a Mellin/Fourier transform in order to put
the kinematical constraints into a factorisable form. So far they are
performed analytically, requiring a dedicated calculation. They cannot easily
cop with the various cuts required by experiments, contrarily to the NLO
calculation implemented with a Monte Carlo method. 
Up to now, the use of 
a NLO Monte Carlo programme in prompt photon phenomenology, supplied by 
motivated scale choices is still legitimate.

\subsection{Isolated photons}\label{isol}

Whereas the contribution from eq.~(\ref{eq:brem}) amounts\footnote{This
statement depends on the choice of scales, especially of $M_{_F}$. The order of
magnitude given here corresponds to a standard choice 
$M_{_F} \sim {\cal O}(p_{_T})$.} to roughly a few tens of percent of the contribution 
from eq. (\ref{eq:sig}) at fixed target energies, it becomes dominant at
colliders at least in the lower $p_{T}$ range. However collider experiments -
besides PHENIX at RHIC - do {\it not} perform inclusive measurements of
photons, strictly speaking. In order to strongly suppress the overwhelming
background of secondary photons coming from the decays of hadrons, mainly
$\pi^{0}$, $\eta$, etc., collider experiments require an isolation criterion on
the photon candidates. A widely used calorimetric criterion, which has the
virtue to be implementable also at the partonic level\footnote{Vetoes on
charged tracks about the direction of the photon are also used experimentaly,
however they cannot be accounted for in a partonic calculation. A detailed
description of the final state including full hadronisation would be
required.}, is the so-called `cone criterion': in a cone about the direction of
the photon defined in rapidity $\eta$ and azimutal angle $\phi$ by
\begin{equation}
\left( \eta - \eta_{\gamma} \right) ^{2} + 
\left( \phi - \phi_{\gamma} \right) ^{2}
\leq R^{2}
\label{crit1}
\end{equation}
the accompanying hadronic transverse energy $E_{T \; had}$ is required to be 
less than some finite amount:
\begin{equation}
E_{T \; had} \leq E_{T \; max} 
\label{crit2}
\end{equation}
$R$ and $E_{T \; max}$ being specified by each experiment, $E_{T \; max}$ being
given either as a fixed value, or as a fixed fraction $\epsilon_{h}$ of the 
photon's $p_{_T}$.

\vspace{0.2cm}

\noindent
Cross sections for producing such isolated photons have been proven
to still fulfill the factorisation property, and are finite to all orders in
perturbation theory for non zero $R$ and $E_{T \; max}$ \cite{Catani:2002ny}. 
Isolation through eqs. (\ref{crit1}, \ref{crit2}) also reduces the 
contribution (F), although it does not kill it completely: a fraction
survives with $z \geq (1 + \epsilon_{h})^{-1}$, which involves the 
{\it same} fragmentations functions $D_{\gamma/k}(z,M_{_F})$ as in the 
unisolated case. The dependence on the isolation parameters $R$ and 
$E_{T \; max}$ is consistently included in the expression describing the hard
subprocess. At colliders energies  the mean value $<z>$ for non isolated photons
from fragmentation is fairly smaller\footnote{For example $<z>$ is roughly 0.7
at the Tevatron, and 0.6 or less at the LHC \cite{ACGG}, whereas typically 
$(1 + \epsilon_{h})^{-1} \geq 0.8 - 0.9$. On the other hand, notice that, at
fixed targets, $<z> \sim 0.9$: in practice photons from fragmentation at fixed
targets are scarcely accompanied by hadrons, they are {\it de facto}
isolated.}  than $(1 + \epsilon_{h})^{-1}$, so that (F) is quite suppressed by
isolation cuts.
Let us stress that $E_{T \; max}$ has to be non zero otherwise 
the calculation of the cross section in perturbative QCD is infrared (IR) 
divergent order by order in perturbation theory, the (D) contribution involving 
a term $\sim \alpha_{S} \, R^{2} \log(p_{T}/E_{T \; max})$. In practice, no IR
sensitivity appears down to fairly low values of $E_{T \; max} \sim$ 1 GeV due
to the smallness of $\alpha_{s}\, R^{2}$. However, the reliability of the
theoretical predicition is jeopardized if the value of $E_{T \; max}$ is nearly
saturated by minimum bias hadrons, thus leaving almost no room for radiation
from the hard event. Another source of trouble for the NLO calculation is
caused by the use of too small a cone size, where the collinear sensitivity
would require an all order resummation of large $\log R$  terms: the NLO
calculation might not be reliable for $R \leq 0.3$ \cite{Catani:2002ny}.

It is important to stress that the isolated cross section, measured
experimentally, cannot be identified with the direct cross section calculated
at the Born level, {\it i.e.} without any contribution
of the fragmentation processes. Indeed, besides the fragmentation piece left
over ($z \geq (1 + \epsilon_{h})^{-1}$) as explained above, higher order terms
originating from the non-collinear fragmentation processes contribute to the
isolated cross sections. Such terms may be important in some kinematical
regions as they correspond to new hard processes, not allowed at the lowest
order : for example, large terms involving the 3-gluon vertex are possible 
at higher orders while they are forbidden at the lowest one.

\subsection{Brief presentation of {\tt JETPHOX}}\label{jetphox}

We have implemented all contributions to (D) and (F) up to NLO in the computer
package {\tt JETPHOX} \cite{cocorico}. This code is a general purpose cross
section integrator of Monte-Carlo type, designed to calculate both single
photon inclusive and photon-jet inclusive cross sections and related
correlations, accounting easily for any kind of experimental cut (e.g. on
kinematics, isolation) implementable at the partonic level. It is the only
available code including both (D) and (F) at NLO in a Monte carlo approach.
Details on the principles and implementation of this code can be found in
\cite{cocorico,Catani:2002ny}. Let us only sketch them briefly.

\vspace{0.2cm}

\noindent 
The treatment of the infrared (IR) soft and collinear singularities of the
partonic transition matrix element combines the phase space slicing
\cite{slicing} and subtraction \cite{subtraction} methods. 
The slicing of phase space is designed as follows. 
For a generic partonic subprocess $1 + 2 \rightarrow 3 + 4 + 5$ two 
outgoing partons, say 3 and 4, have a high $p_{T}$ and are well separated in
phase space, while 5, say, can be soft, or collinear to either of
the four others. The phase space is sliced using two arbitrary (unphysical)
parameters $p_{Tm}$ and $R_{Th}$, with $p_{Tm} \ll ||\vec{p}_{T\,3,4}||$ 
and $R_{Th} \ll 1$, in four parts:

\begin{itemize}
\item[-] Part I corresponds to $||\vec{p}_{T\,5}|| < p_{Tm}$. This
cylinder supports the IR and initial state collinear singularities. 
It also yields a small fraction of the final state collinear singularities. 

\item[-] Part II a corresponds to $||\vec{p}_{T\,5}|| > p_{Tm}$, 
$\vec{p}_{T\,5} \in C_3$, 
where $C_3$ is 
the cone 
%about the direction of $3$
defined by $(y_5-y_3)^2+(\phi_5-\phi_3)^2~\leq~R_{th}^2$. 
It supports the final state collinear singularities when 
$5$ is collinear to $3$.

\item[-] Part II b is defined in a similar way as II a but with 
the replacement of $3$ by $4$. It supports the final state collinear 
singularities when $5$ is collinear to $4$.

\item[-] Part II c is the remaining region:
$||\vec{p}_{T\,5}|| \geq p_{Tm}$, and $\vec{p}_{T\,5} \not{\!\!\in} \, C_3$, 
$C_4$.
\end{itemize}

\noindent
Collinear and soft singularities, which appear on parts I, II a and II b, are
first regularised by dimensional continuation from $4$ to $n = 4 - 2
\varepsilon$ with $\varepsilon < 0$. Then, the $n$-dimensional integration
over particle 5 is performed analytically over these parts. After combination
with the corresponding virtual contributions, the infrared singularities
cancel, and the remaining collinear singularities which do not cancel are
factorised and absorbed in the parton distribution functions or fragmentation
functions. The resulting quantities correspond to pseudo cross sections in 
which the the ``integrated out" parton 5 is unresolved from the remaining four hard
partons. The word ``pseudo" means that they are not genuine cross sections,
namely they are not necessarily positive, and they depend on the arbitrary
choice of factorisation scheme\footnote{The $\overline{MS}$ factorisation
scheme is used.}. Part II c yields no singularity, and is thus treated
directly in $4$ space-time dimensions. These pseudo cross sections, as well
as the transition matrix elements on the part II c, are then used to sample
partonic events according to a Monte-Carlo method. Last, these partonic events
are projected onto histograms thus providing any desired distribution. 

\vspace{0.3cm}

\noindent
By virtue of the factorisation theorem, the contribution (F) alone also 
provides the NLO cross sections for inclusive hadron- and associated hadron + 
jet production, once the parton-to-photon fragmentation functions have been 
replaced by fragmentation functions of partons to the hadron species considered.
The phenomenology of correlations in associated prompt photon + jet and 
hadron + jet production using {\tt JETPHOX} will be presented in a future
article \cite{correlations}.

\section{Comparison with data}
\label{comp-data} 

All the comparisons between data and NLO calculations provided in this section 
are made using the parton distribution functions of set CTEQ6M~\cite{cteq6m} 
($\alpha_s(M_Z) = .118$)
and the parton-to-photon fragmentation functions of set 
BFGW (set II)~\cite{BFG}.
Whenever the scales $\mu_{R}$, $M$ and $M_{_F}$are given a common value, the 
latter is noted $\mu$. The $\overline{MS}$ scheme is used throughout.

\subsection{New data from D$\emptyset$ and PHENIX}

%%%%%%%\subsection{New D$\emptyset$ data in isolated photons}

\begin{figure}[htb]
\centering
\includegraphics[width=4in,height=3in]{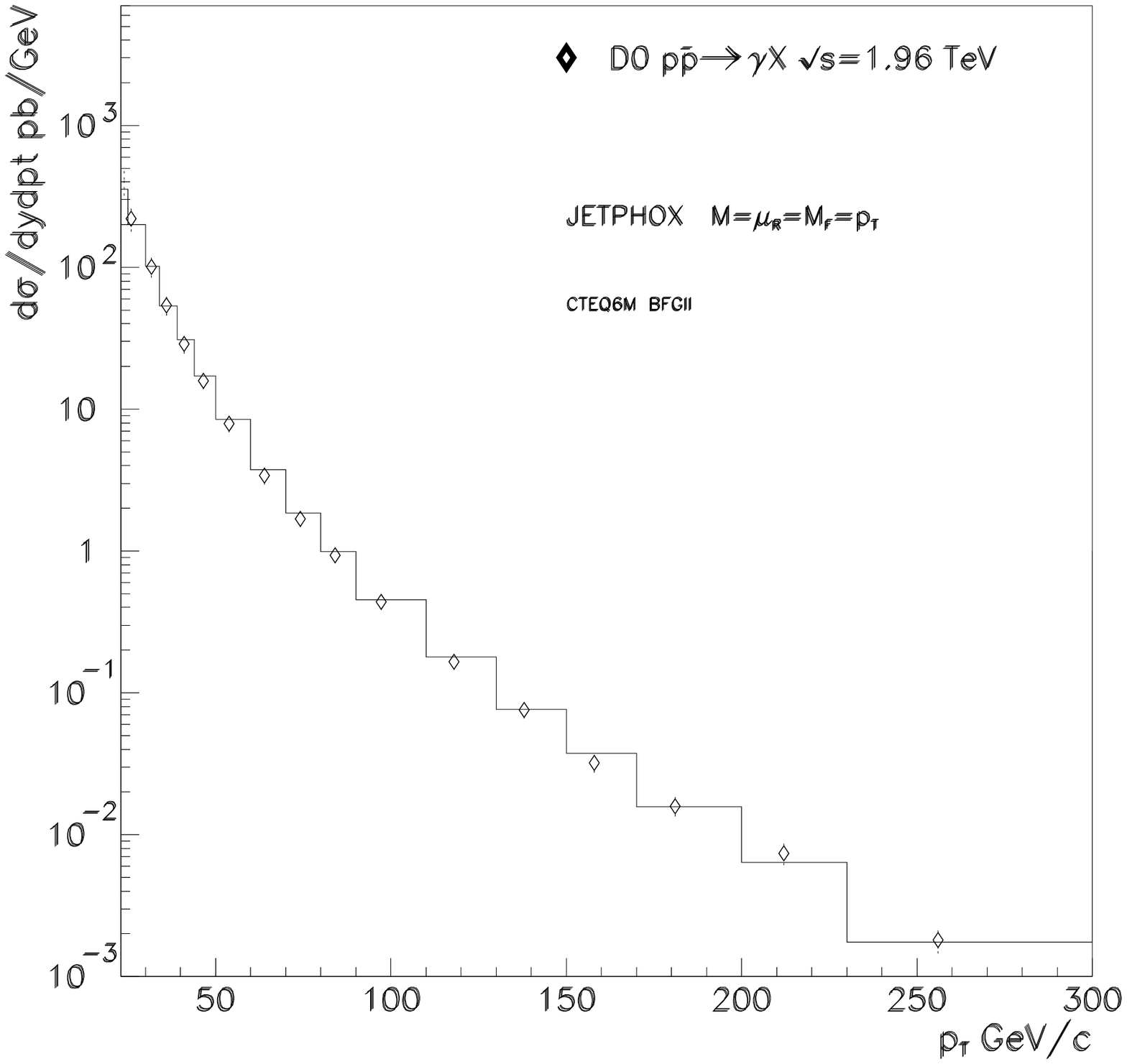}
\caption{The isolated D$\emptyset$ photon cross section in the 
central ($|\eta | < .9$) pseudorapidity region. The histogram is the NLO 
QCD prediction discussed in the text. The errors are the sum of the
statistical and systematic errors. The scales are $\mu = p_T$.}
\label{fig:1}
\vspace{1.cm}
\includegraphics[width=4in,height=3in]{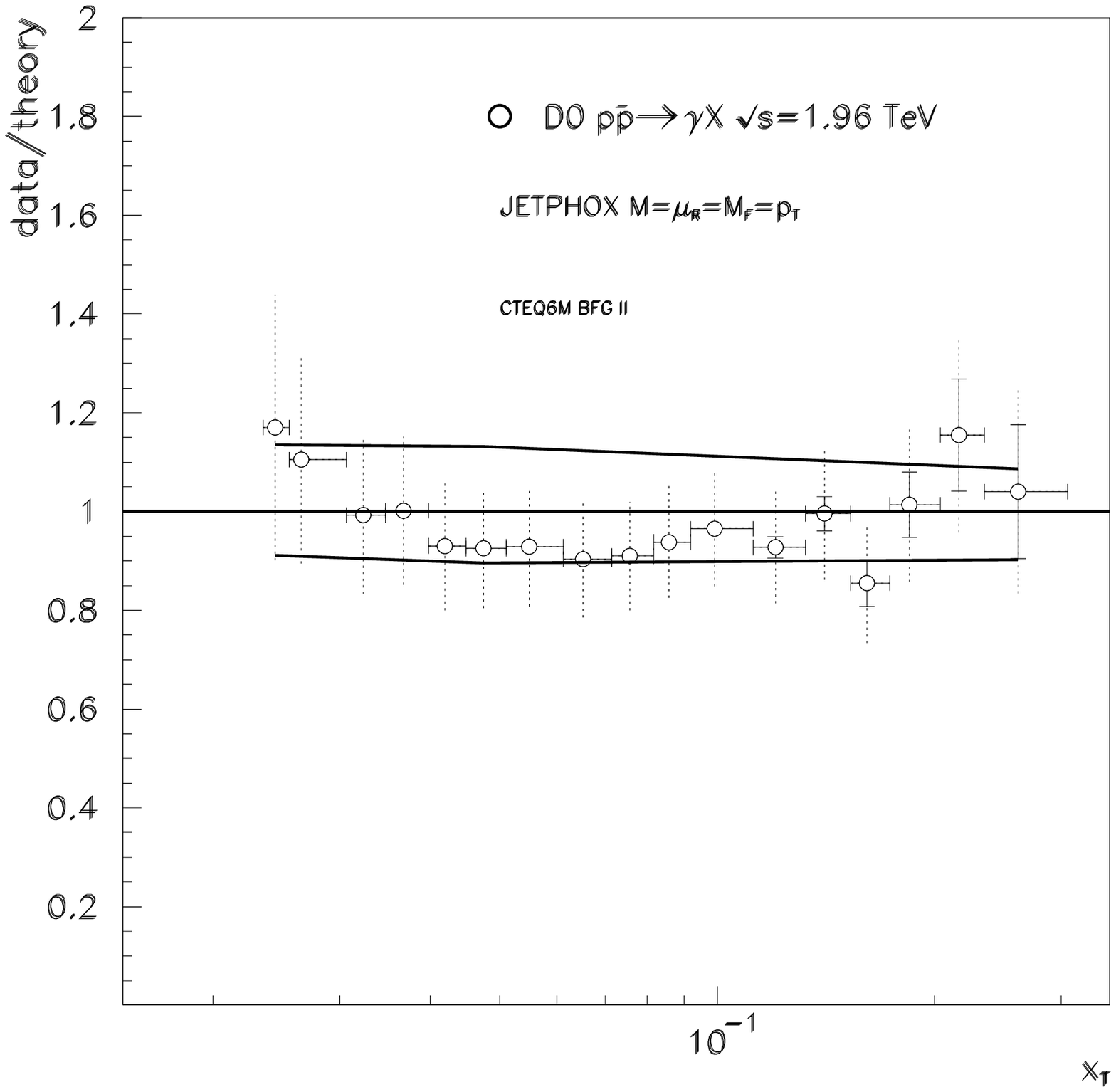}
\caption{The ratio of D$\emptyset$ data to NLO QCD obtained with $\mu = p_T$. 
Ratios of the predictions for $\mu = p_T/2$ ($\mu = 2 p_T$) to the nominal 
theory ($\mu = p_T$) are shown by the upper curve (lower curve).}
\label{fig:2}
\end{figure}

We start this section by the analysis of the new data taken by
the D$\emptyset$ collaboration \cite{Abazov:2005wc} during the Tevatron Run II at $\sqrt{s}
= 1.96$~TeV. The measured cross section, in the range 23 GeV $< p_T <$ 300 GeV,
is for isolated photons and we account for the D$\emptyset$ isolation criterion by
requiring that the hadronic transverse energy measured in a cone of radius $R
= \sqrt{\Delta\phi^2 + \Delta \eta^2} = 0.4$ around the photon is smaller than
10\% of the photon's $p_T$. Theory and data are compared in Fig.~\ref{fig:1} and
\ref{fig:2}, the theoretical curves being obtained with the inputs specified 
at the beginning of this section. 
\noindent
In Fig.~\ref{fig:1}, all the scales have been set equal to $p_T$. We see that 
the agreement between data and NLO QCD is excellent in the whole $p_T$-range in
which the cross section falls by about five orders of magnitude. A more precise
comparison is shown in Fig.~\ref{fig:2} for the ratio data/theory calculated 
with several choices of scales between $p_T/2$ and $2p_T$. The sensitivity to 
the changes in the common scale $\mu$ is of some $\pm 10$ \% in the whole 
$p_T$-range. Using MRST 2004 \cite{martin:2004ir} instead of CTEQ6M changes the predictions by 
$\pm 2$ \%. The present experimental errors have the size of the variations 
coming from the scale changes; with this accuracy there is no evidence of any 
systematic deviation of the theory with respect to data.

%%%%%%%%\subsubsection{New PHENIX data on both inclusive and isolated photons}

Another set of recent - still `very preliminary' - data is presented by the 
PHENIX collaboration \cite{Adler:2005,Okada:2005}
at RHIC. Measurement are done at $\sqrt{s} = 200$~GeV, an intermediate energy
between collider measurements at $\sqrt{s} = 630$~GeV and fixed targets
measurement at $\sqrt{s} \leq 40$~GeV which will be discussed below. 
These data cover the range $4 < p_T < 17$~GeV and correspond to two methods of 
analysis: a subtraction method in which the $\pi^0$ background is identified 
and subtracted (inclusive prompt photon cross section), and an isolation method.
We start with a discussion of the isolated data.

\begin{figure}[htb]
\centering
\includegraphics[width=4in,height=3in]{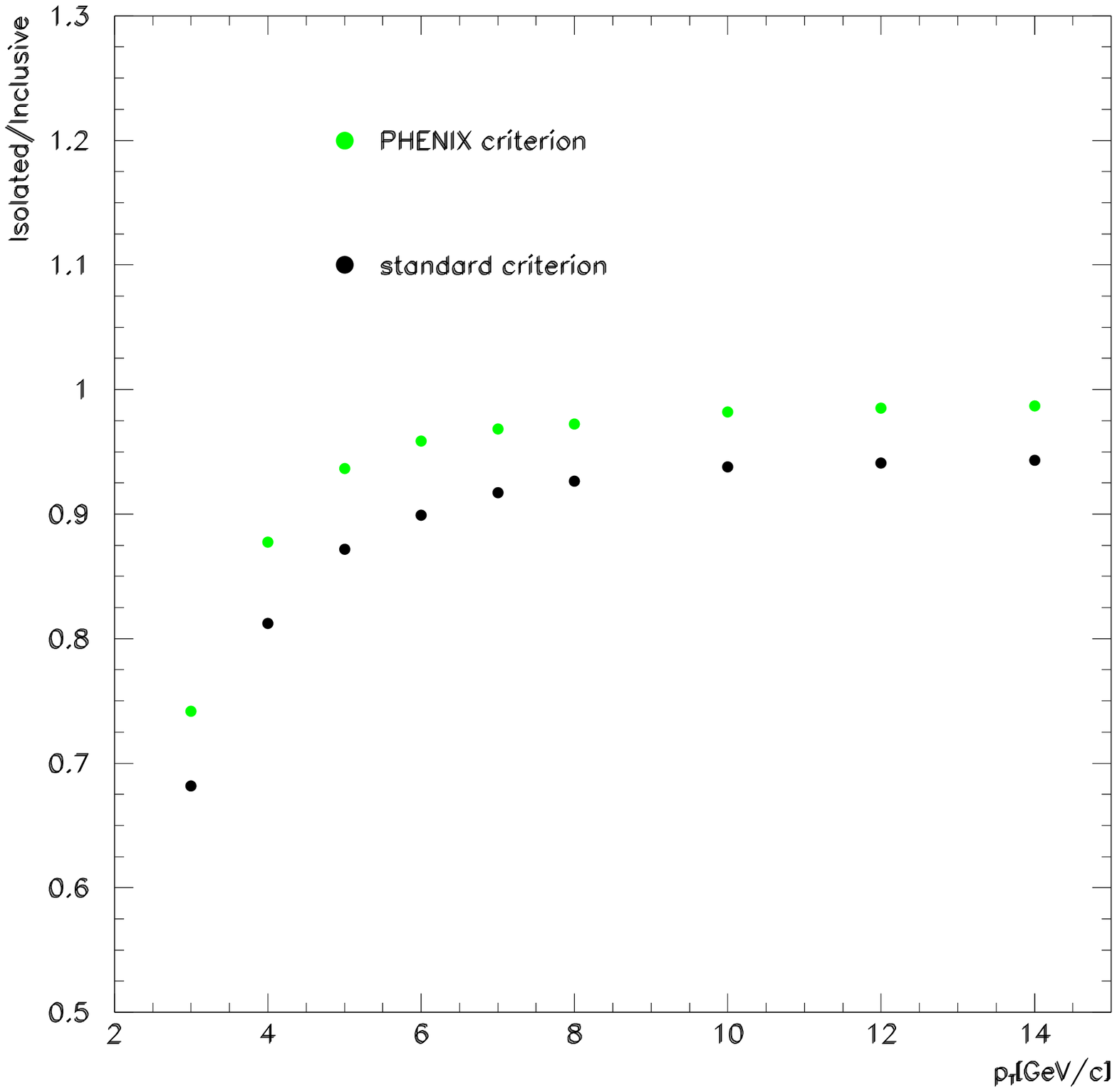}
\caption{Ratio $\sigma^{th \, isol}/\sigma^{th \, incl}$
of the NLO isolated cross sections to the NLO inclusive prompt-photon cross sections 
at $\sqrt{s} = 200$~GeV.
The scales used are $\mu= p_T/2$. PHENIX criterion: colored dots; standard
criterion: black dots}
\label{fig:isol-phenix}
\end{figure}

The isolation criterion used by the PHENIX collaboration is fitted to the
acceptance of its detector. Photons are detected in the coverage $-.30 \leq
\eta \leq .30$ and $-.73 \leq \phi \leq .73$ and the hadronic transverse energy
is measured in a cone of radius $.5$ if it also falls into the region $-.35
\leq \eta \leq .35$, $-\pi/4 \leq \phi \leq \pi /4$. When the hadronic energy
is outside the acceptance, it is not taken into account. The fraction of
hadronic energy thus observed should be less than 10 \% of the photon momentum.
This criterion amounts effectively to implementing isolation in a smaller
region about the photon and we expect the effect of isolation to be smaller
than the one due to the standard procedure when all the energy in the cone is
taken into account. The NLO code JETPHOX allows us to study the effect of the
PHENIX isolation compared to the standard isolation. The ratios of isolated
cross sections over the inclusive one are shown in Fig.~\ref{fig:isol-phenix}
where the scale $p_T/2$ is used for the theoretical predictions. First, let us
note that the isolation effect is large at low $p_T$. At larger values of
$p_T$  the average value of $z$ increases and the isolation cut is less
effective. Also we note that the PHENIX criterion and the standard one lead to
appreciable differences in the predictions, of the order of $10$ \%  at low 
$p_T$.

Note that the NLO calculations are performed at the parton level for the QCD
hard process and do not account for hadronisation effects which can be large
at low pt. Moreover we do not describe the soft underlying event with
transverse energy which can also fall into the isolation cone. This
contribution may even cut the Born contribution (unaccompanied photon). This
effect has been studied by the H1 collaboration in the photoproduction of
prompt photons and found to be non negligible~\cite{h1-2004}.

\begin{figure}[htb]
%\vspace{9pt}
\centering
\includegraphics[width=4in,height=3in]{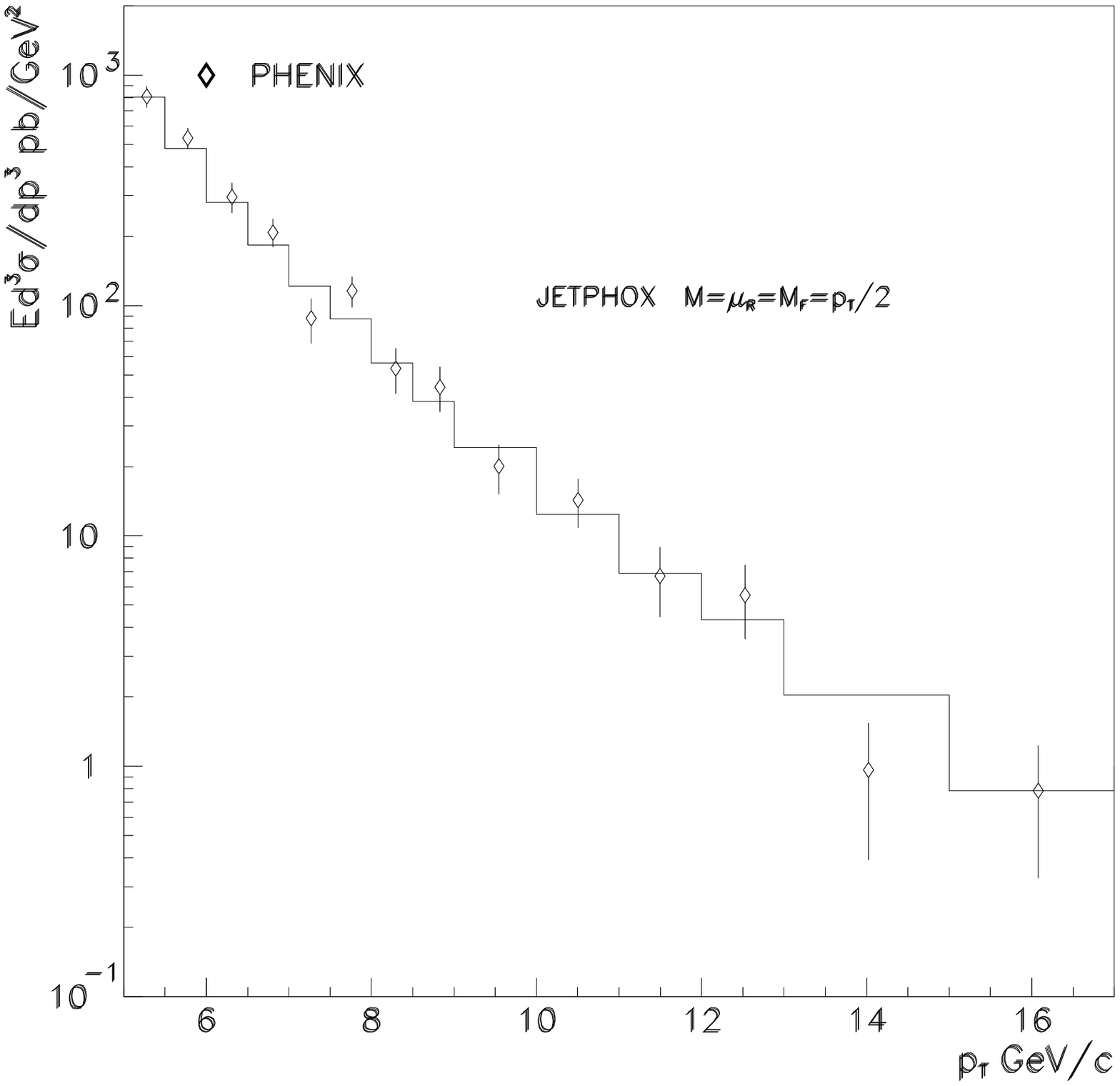}
\caption{The PHENIX isolated prompt-photon cross section at 
$\sqrt{s} = 200$~GeV compared with NLO cross section predictions. 
Only the statistical errors on the data are shown. The scales used
in the theoretical calculation are $\mu= p_T/2$.}
\label{fig:isol-section}
\end{figure}
\noindent

\begin{figure}[htb]
\centering
\includegraphics[width=4in,height=3in]{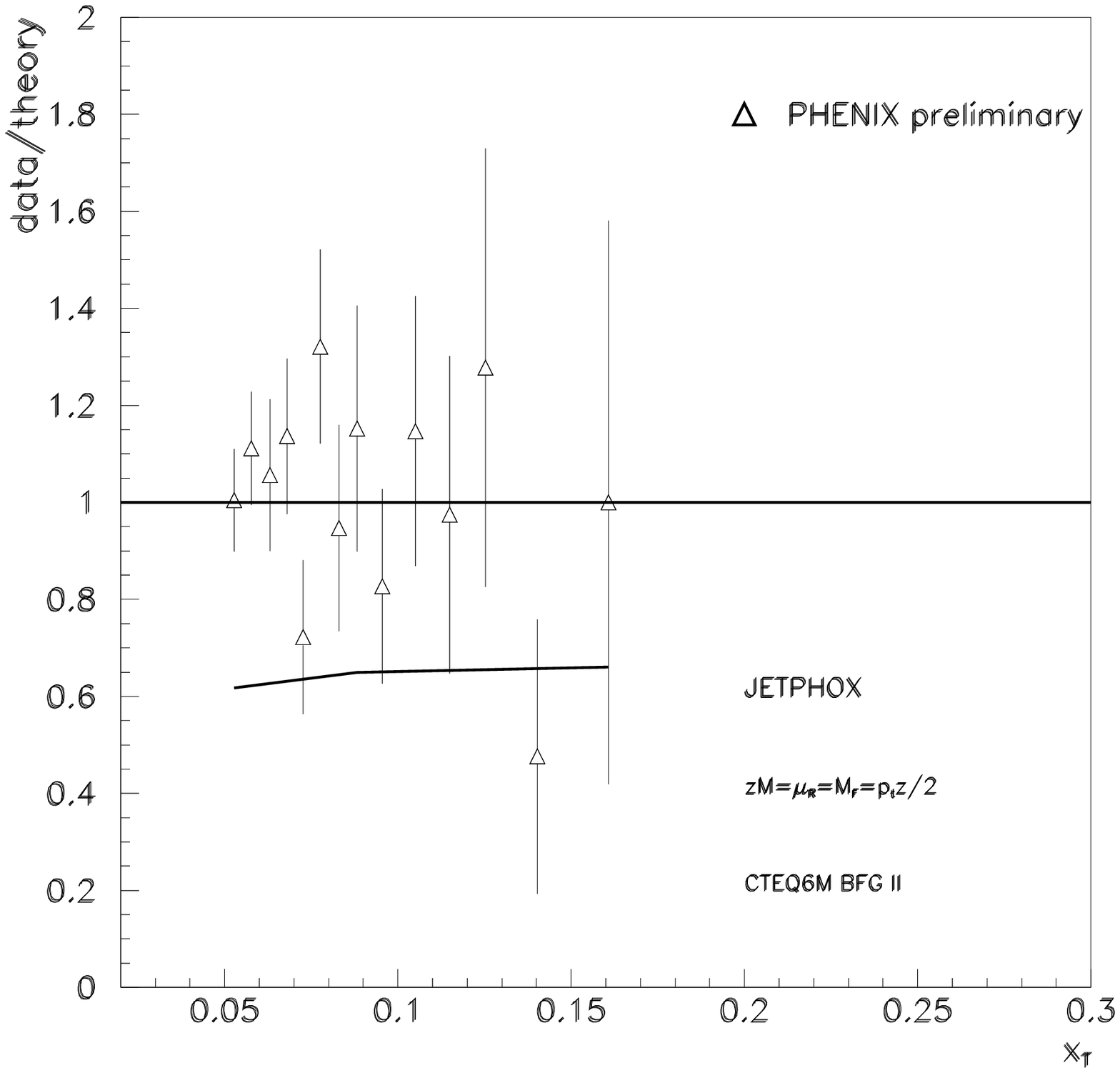}
\caption{The ratio of PHENIX isolated photon data to NLO QCD using 
$\mu = p_{_T}/2$. 
The lower curve corresponds to the ratio 
$\mbox{NLO}(2 p_T)/\mbox{NLO}(p_T/2)$. Only the statistical errors on the data
are shown.}
\label{fig:isol-ratio-phenix}
\end{figure}

The comparison of the  PHENIX isolated cross section with the NLO QCD
prediction with the scale $\mu = p_{T}/2$ is  shown in
Fig.~\ref{fig:isol-section}.  The theory agrees very well with data within
errors (systematic errors are not shown in Fig.~\ref{fig:isol-section}). A more detailed comparison is
performed in Fig.~\ref{fig:isol-ratio-phenix} in
terms of ratios  data/theory calculated for two different choices of scales.
The statistical errors are relatively large, of the order of the theoretical
uncertainty when varying the common scale from  $p_T/2$ to $2p_T$. However, the
standard choice  $p_T/2$ reproduces the data extremely well over the whole $p_T$
range in which the cross section varies by  a factor $10^3$.

\begin{figure*}[htb]
%%%%%%%%%%%%%%%%%%\vspace{9pt}
\centering
\includegraphics[width=6in,height=4in]{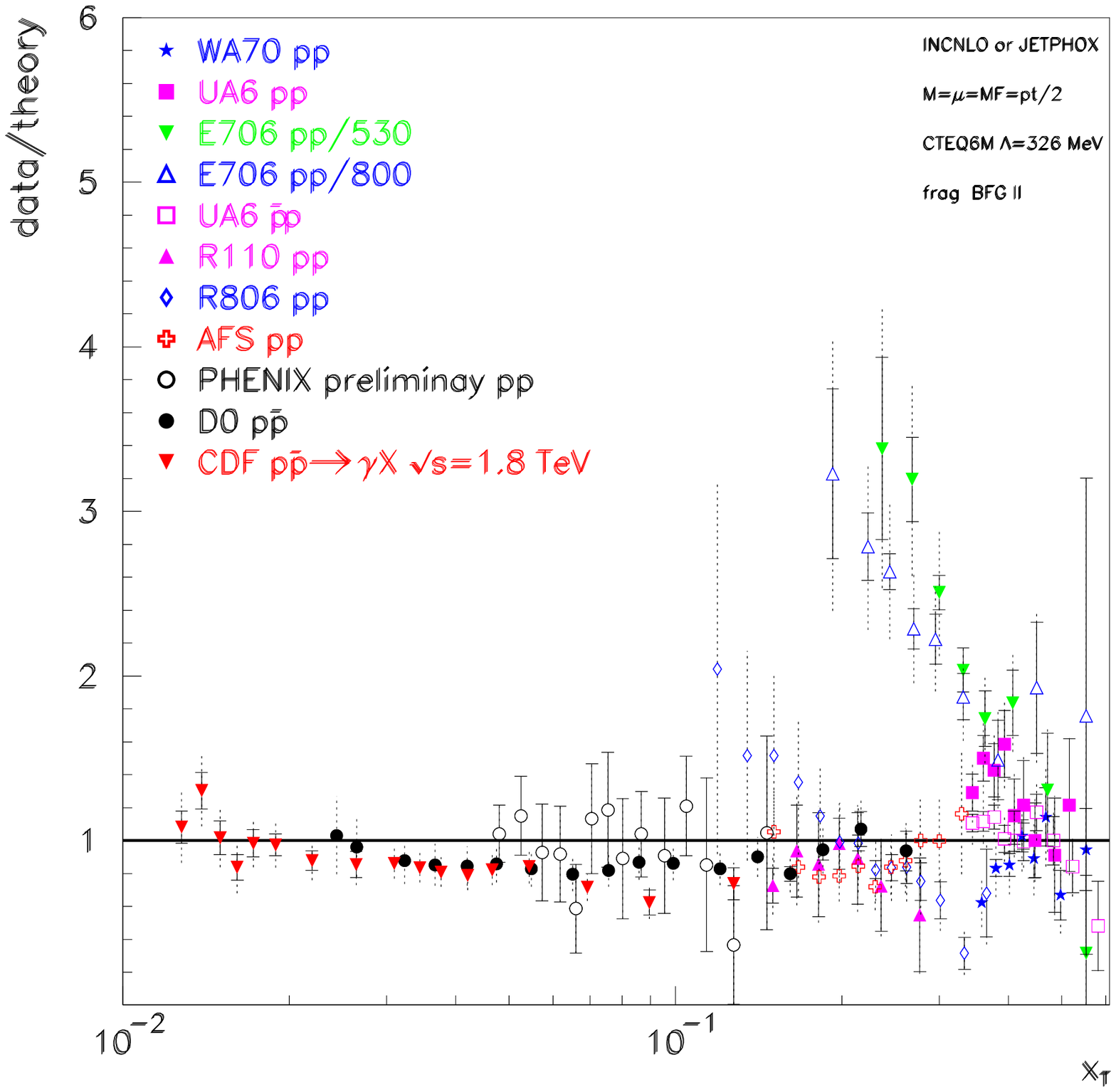}
\caption{Ratios data/theory for collider and fixed target data with the 
scale $\mu = p_T/2$. For PHENIX and lower energy data the inclusive cross
section is used while the isolated one is used for CDF and D$\emptyset$.
Statistical erros only for PHENIX data.}
\label{fig:isol-ratio-all}
%%%%%%%%%%%%%%%\end{figure*}
%%%%%%%%%%%%%%%\begin{figure*}[htb]
\vspace{9pt} 
%%%%%%%%%%%%%%%\centering 
\includegraphics[width=6in,height=4in]{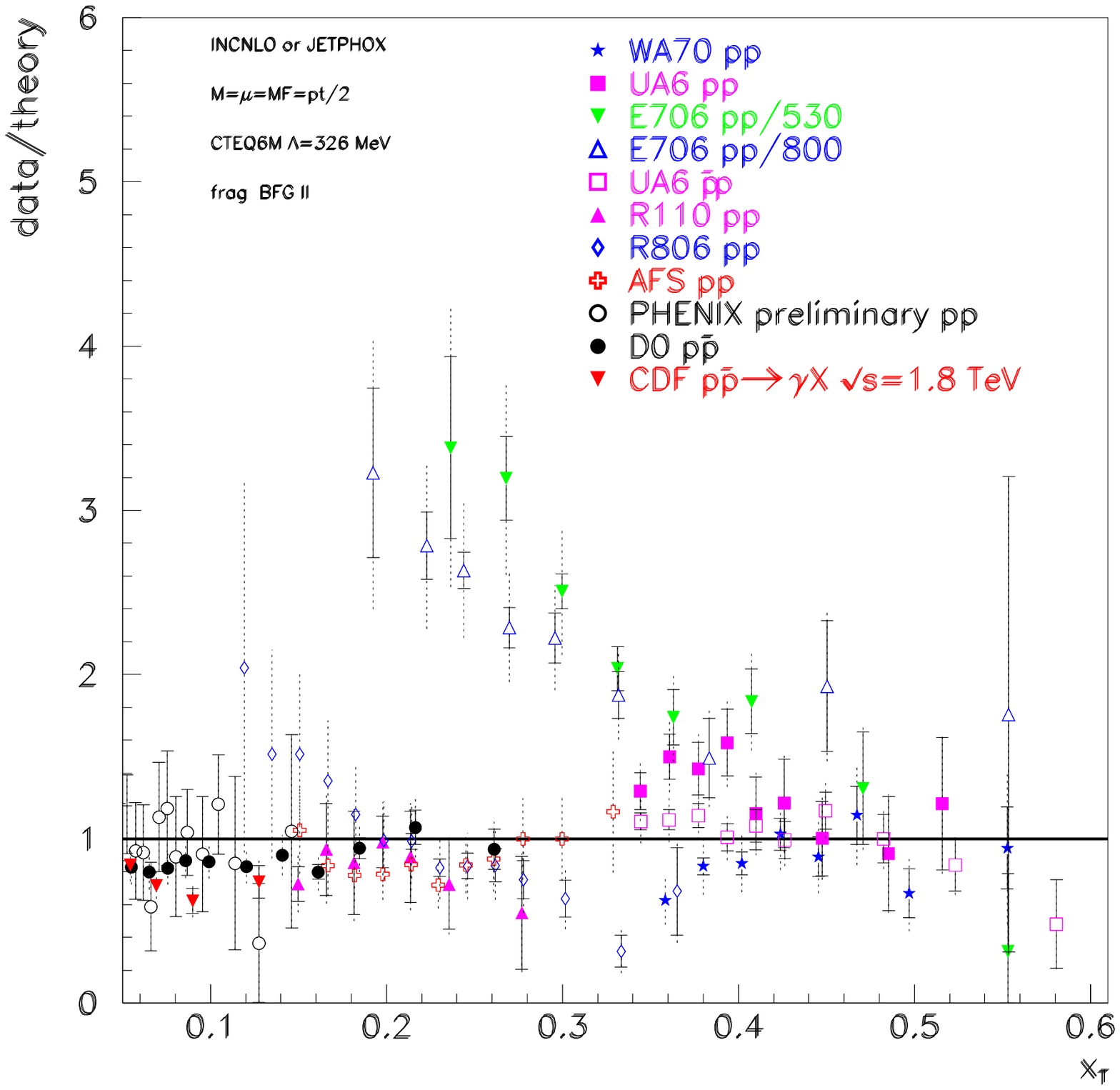}
\caption{Zoom on the large  $x_T$ data with a linear scale in $x_T$.
%%%%%%%Same as Fig.~\ref{fig:isol-ratio-all}, but with a linear scale in $x_T$.
} 
\label{fig:isol-ratio-all-lin}
\end{figure*}

\subsection{Previous world data in the light of the new data from 
D$\emptyset$ and PHENIX}

As there is some overlap in $x_T$ between these
new data and some of the previous ones, in particular in the controversial
0.2 to 0.3 range previously covered by the ISR and E706 experiments, it is interesting to
reconsider how the  `world data' is described by theory.

%%%%%%%%%\subsubsection{Fixed targets data on inclusive (un-isolated) photons}

We consider now the inclusive $pp$ and $p\overline{p}$ data coming from the 
fixed target experiments from WA70 ($\sqrt{s} = 23$~GeV) \cite{WA70}, UA6 
($\sqrt{s}= 24.3$~GeV)  \cite{UA6}, E706 ($\sqrt{s} = 31.6$~GeV and 38.8~GeV)
\cite{E70604}, from the ISR experiments  ($\sqrt{s} =63$~GeV) R110 \cite{R110},
R806 \cite{R806}, AFS \cite{R807} and the isolated data from CdF  at $\sqrt{s}
= 1.8$~TeV \cite{CDF2002,CDF2004}. The collider data at $\sqrt{s} = 630$~GeV 
will be discussed separately. Below we shall also compare the {\it pBe}  data
from  the E706 experiment \cite{E706} with the more recent $pp$ data of the
same  experiment~\cite{E70604}. The comparison is done with the scales $\mu = p
_T /2$ in terms of ratios data/theory. As explained in section 2.1, such a
scale  is motivated for the fixed target and ISR range by the recent resummed
calculations \cite{los,cmn,cmnov,sv,deFlorian:2005wf}. 
The results are shown  in
Fig.~\ref{fig:isol-ratio-all} which exhibits the striking agreement between
theory and data in the  whole $x_T$ range, with the exception of the E706
data\footnote{Note that the  smallest $x_T$-values of the E706 data correspond
to values of $p_T$ down to 3.5  GeV/c (averaged $p_T \simeq 3.73$~GeV in the
first bin).}. This last point has been already discussed  at length for {\it
pBe} data in ref.~\cite{afgkpw}. Here the new features are the new D$\emptyset$
and  PHENIX data  which confirms the `world' agreement between theory and data.
We emphazise the very good agreement between theory and the PHENIX inclusive
data which confirms that already shown with the isolated data in 
Figs.~\ref{fig:isol-section} and \ref{fig:isol-ratio-phenix}.
If we disregard the E706 data, there is no evidence for any systematic
discrepancy between data and theory. In Fig.~\ref{fig:isol-ratio-all}, the fixed target data
are somewhat squeezed by the logarithmic scale. In Fig.~\ref{fig:isol-ratio-all-lin} we
emphasized the fixed target domain by using a linear scale which makes the
discrepancy between the E706 data and the other data more obvious. 
The $x_T$ range from 0.15 to 0.3 is well described all the way from TEVATRON collider data 
at $\sqrt{s}=1.96$~TeV down to ISR data at $\sqrt{s}=63$~GeV. The disagreement
cannot be due to the 
`low' center of mass energy, ($\sqrt{s}=31.6,\ 38.8$~GeV), of the E706
experiment  since the WA70 and UA6 fixed target data at even lower energies, 
$\sqrt{s}=23$ and $24.3$~GeV respectively, (and higher $x_T$) are in good
agreement with theory. Let us note that a reasonably 
accurate determination of $\alpha_s$, in good agreement with  determinations
from other processes,was performed by the UA6
collaboration~\cite{Werlen:1999ij}.    
 
%%%%%%%%%%%\subsection{Collider data on isolated photons} 

\begin{figure*}[htb]
%\vspace{9pt}
\centering
\includegraphics[width=6in,height=4in]{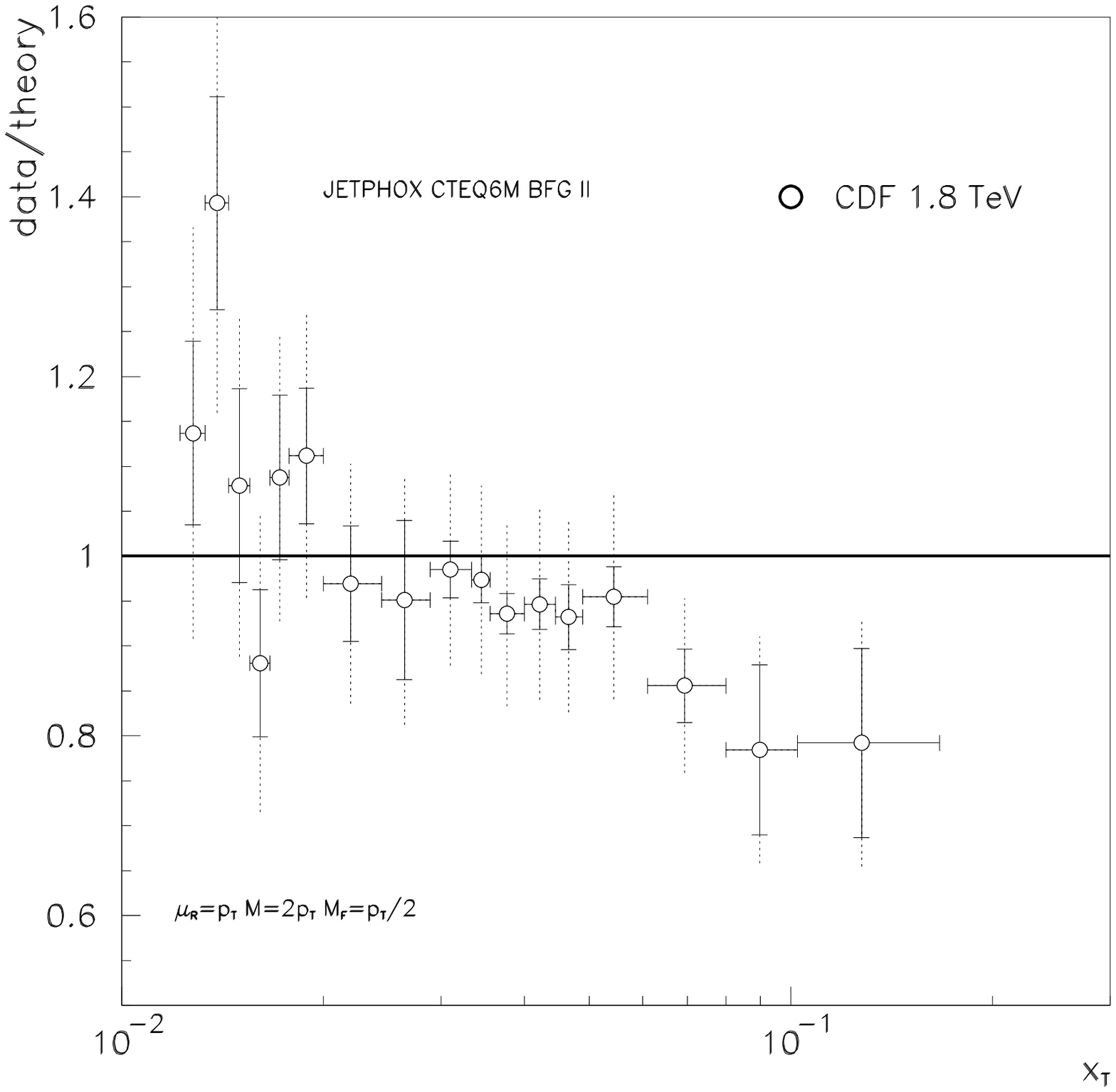}
\caption{CDF data versus theory for the choice of scales as shown in the 
figure.}
\label{fig:ratio-cdf}
\end{figure*}

Let us continue this `world' comparison by a comment on the low-$x_T$ part of 
the CDF data. It has been often claimed 
\cite{HKKLOT,Baur:2000xd,CDF2002,CDF2004}   that there is  a
disagreement between data and theory, the latter being unable to explain the 
rise of the former, a 20 \% effect in the ratios of
Fig.~\ref{fig:isol-ratio-all} and \ref{fig:isol-ratio-all-lin}. Here we would
like to point out  that the significance of this rise is much reduced when
experimental errors  and theoretical uncertainties are  taken into account.
Concerning this last point, the slope in the ratio data/theory depends on the
choice of the scales as already noted in~\cite{vv}. An example is given  in
Fig.~\ref{fig:ratio-cdf} where the ratio data/theory is shown for the scales
$\mu_R = p_T$, $M = 2p_T$, $M_F = p_T/2$. The choice yields a slight flattening
of the curve.  

\begin{figure*}[htb]
\centering
\includegraphics[width=6in,height=4in]{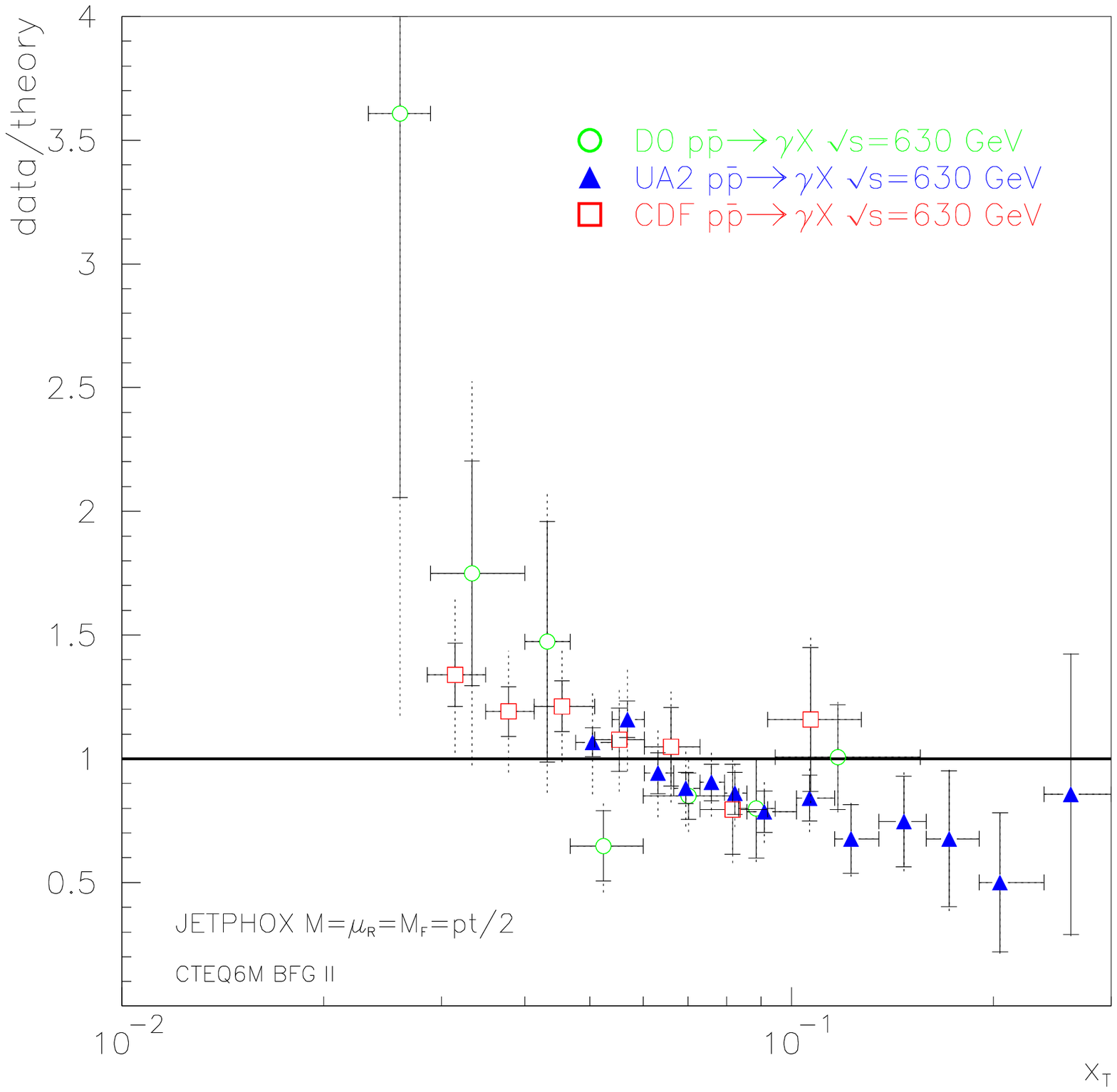}
\caption{Ratios data/theory for collider data at $\sqrt{s} = 630$~GeV. 
The scales are  $\mu = p_T/2$.}
\label{fig:ratio-630}
\end{figure*}

The comparison with the isolated collider data at $\sqrt{s} = 630$~GeV from
UA2  \cite{UA2}, CdF \cite{CDF2002} and D$\emptyset$ \cite{D02001} are shown in
Fig.~\ref{fig:ratio-630}. The  errors being rather large, it is difficult to draw
any precise conclusion from  these results. But within the errors, the
agreement data versus theory is good : the CDF data tend to be systematically
somewhat above the predictions while the UA2 results tend to be slightly below.
The apparent slope effect of the D$\emptyset$ data, at the lowest values of
$x_T$, is not meaningful when taking into account the large systematic errors.
As for UA1 data~\cite{UA1}, the large error bars do not constrain the theory 
very much, and we do not show these data here. One can note however that the
corresponding ratios would be above one.
\begin{figure*}[htb]
\centering
\includegraphics[width=6in,height=4in]{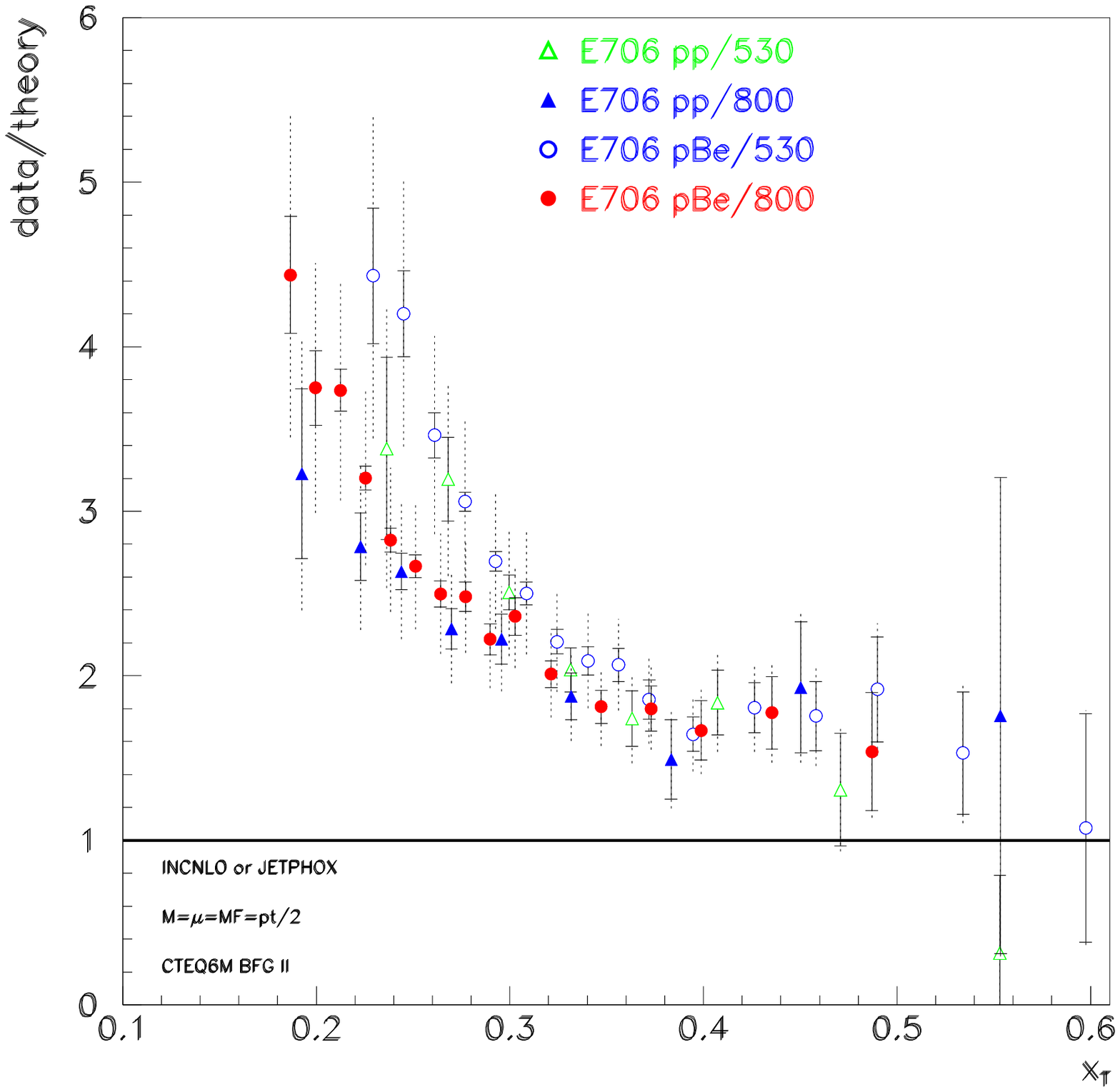}
\caption{Comparison of $p p$ and $p Be$ data of E706 normalised by the
theoretical predictions with scales $p_T/2$.}
\label{fig:ratio-E706}
\end{figure*}

We now turn to a comparison of the recent $pp$ data from the E706 
collaboration \cite{E70604} with the $pBe$ data \cite{E706} of the same 
collaboration. This is done in Fig.~\ref{fig:ratio-E706}. We clearly
distinguish two domains for the E706 data, one for  $x_T \, \geq \,  .34$ and a
second one for lower values of $x_T$.  In the large-$x_T$ domain, the ratio
data/theory is approximately flat, or  slightly decreasing with an average
value close to 1.8. Proton-proton data  and proton-Berylium data are compatible
within errors.  The problem met here by theory is that  of the normalisation.
Even the use of a small scale $\mu = p_T/3$ does not  reconcile NLO QCD with
data \cite{afgkpw}. The low-$x_T$ domain is characterized  by a marked rise of
the ratios, up to 4.5 (at $p_T \simeq 3.73$~GeV) when $x_T$ decreases.  Here
also $pp$ and $pBe$ data are compatible within errors. NLO QCD cannot  explain
this rise and another mechanism, the $\kappa_T$-enhancement, has  been put
forward by the E706 collaboration to explain the shape and  normalisation of
the cross sections. Let us note that no other experiment  needs such an
enhancement.

\begin{figure*}[htb]
\centering
\includegraphics[width=7.in,height=8.in]{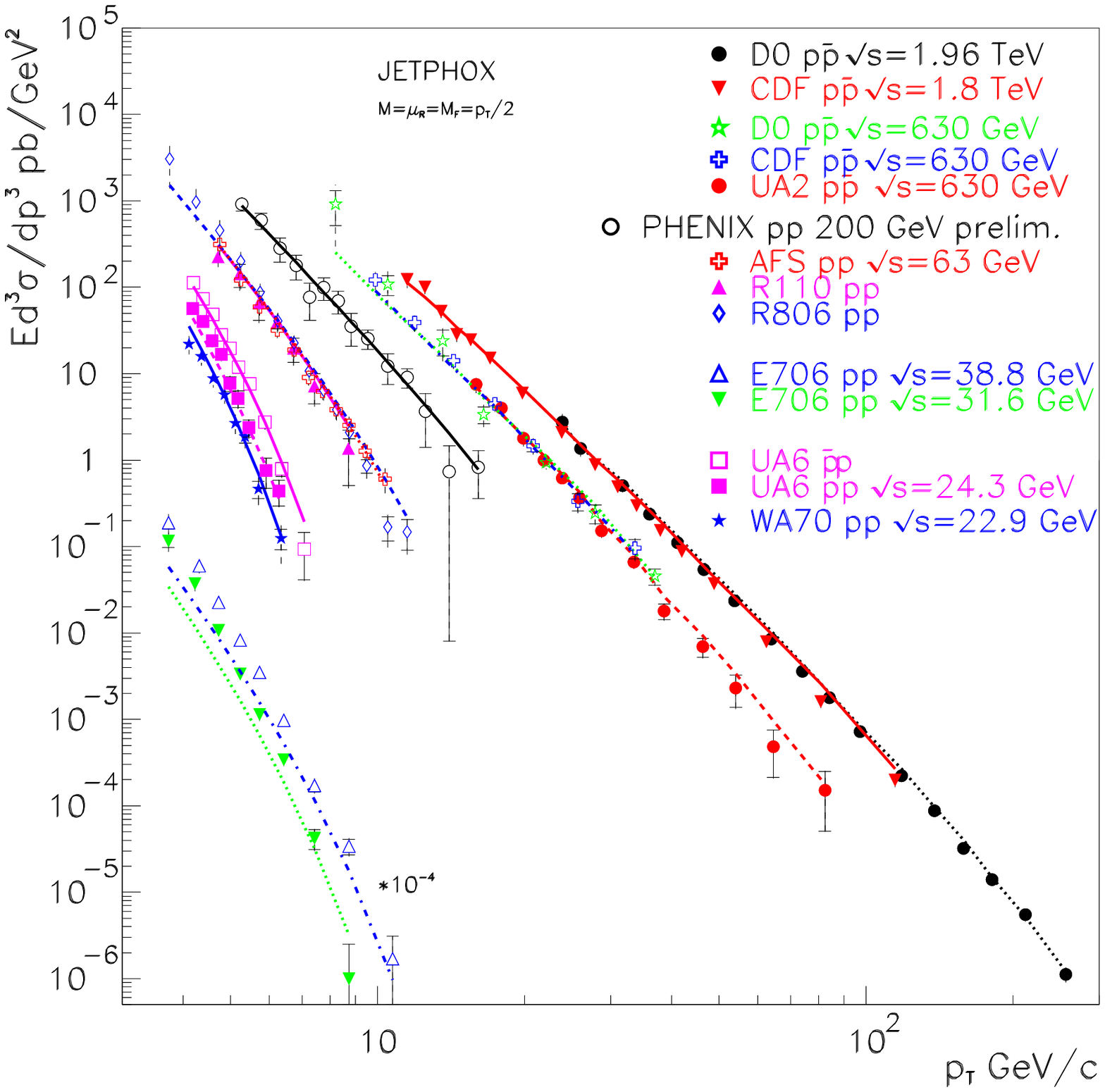}
\caption{World's inclusive and
isolated direct photon productions cross sections measured in proton-proton and
antiproton-proton collisions compared to JETPHOX NLO predictions using BFG II
(CTEQ6M) for fragmentation (structure) functions and a common scale $p_T/2$. For
the clarity of the figure the E706 data are scaled by a factor $10^{-4}$.}
\label{fig:all}
\end{figure*}

\begin{figure}[htb]
\centering
\includegraphics[width=5in,height=5in]{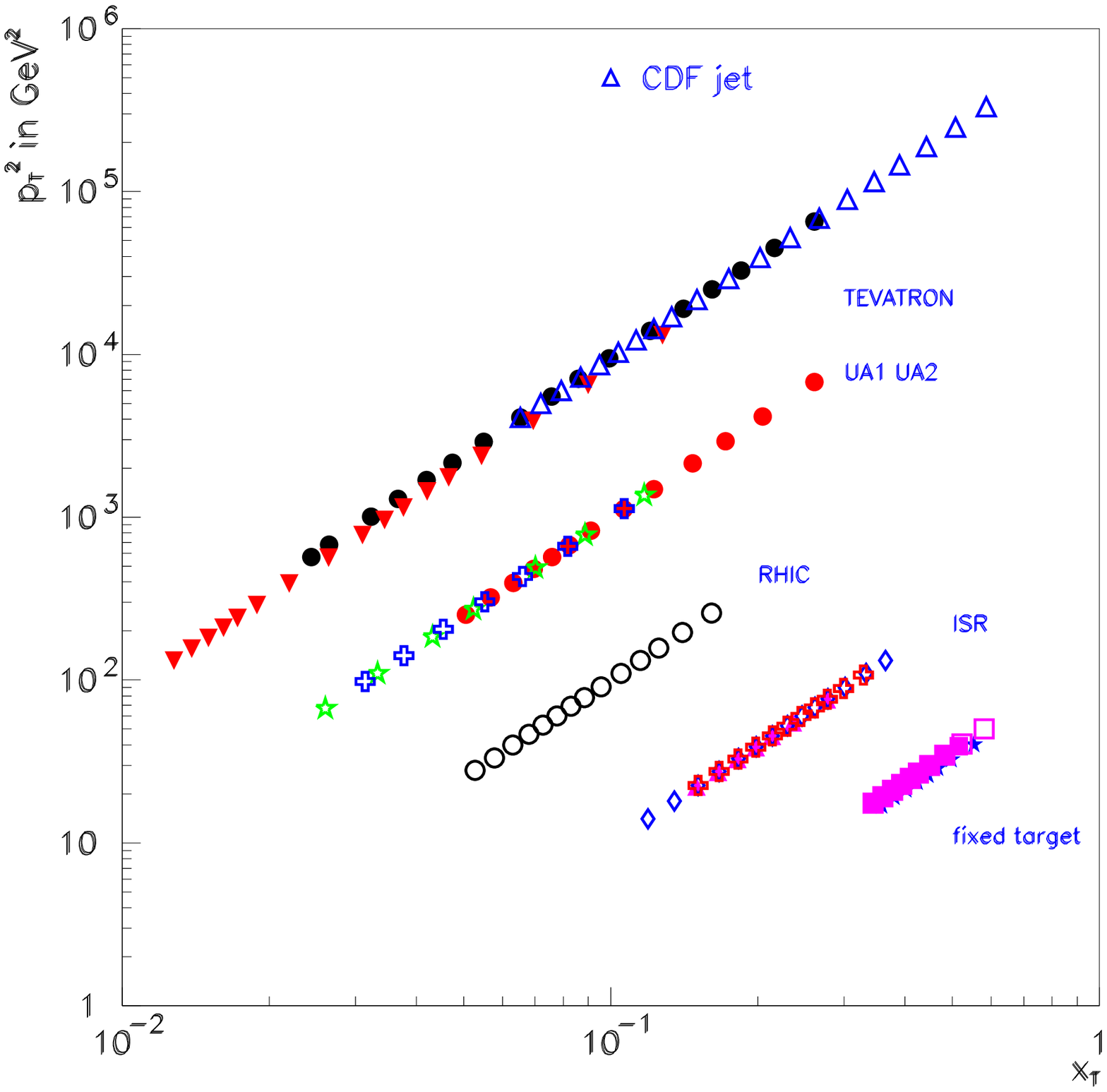}
\caption{The kinematical region probed by prompt photon experiments compared to
that relevant for jet production. Each data point is represented by a symbol
as in Fig. 11 for photons, and by open triangles for jets.}
\label{fig:kinem}
\end{figure}

We end this section by collecting all available prompt photon cross sections $E
d^3\sigma / d p^3$  (inclusive or isolated) on the same plot and comparing them
with theoretical predictions evaluated with the scale $\mu = p_T/2$
(Fig.~\ref{fig:all}).  The data span two orders of magnitude in energy and 
there is agreement over nine orders of magnitude in the cross sections between
theory and experiments. This is comparable to the agreement between theory and
D$\emptyset$ run 2 data for the jet cross section~\cite{D0jet}  and similarly
for CDF data~\cite{cdfjet}. However the prompt photon data and the jet data do
not cover the same kinematical region defined in the $(x_T = 2 p_T/\sqrt{s}, \
p_T^2)$ plane (which is the equivalent, for large $p_T$ processes, of the $(x,\
Q^2)$ plane of deep inelastic scattering) as shown in Fig.~\ref{fig:kinem}. The
combined data therefore give an extremely strong test of QCD.

\section{Conclusions and outlook}
\label{conclusions}

Two new direct photon data sets shed a new light on the understanding of such
processes within the QCD framework, and they confirm that the $\sqrt{s}$
dependence of the reaction can be properly described within the NLO formalism. 
Agreement between data and theory from $\sqrt{s} = 23$~GeV to
1.8~TeV is very good over nine orders of magnitude in the cross section.

This is in contrast to the view, based only on data from the E706
experiment, that the direct photon data cannot be  understood in the NLO QCD
framework without resorting to a non perturbative `$k_T$ kick'. Indeed, although
refinements like resummed calculations have been important to reduce the
theoretical uncertainties, they have not permitted to reconcile one data set
(namely E706) with theory without using large non perturbative parameters. Such
parameters are not needed by other experiments.

\section*{Acknowledgments}

The authors thank Dmitry Bandurin, Gerry Bunce, Joe Huston and Kensuke Okada
for interesting correspondance and discussions.


\begin{thebibliography}{99}
%

\bibitem{Adler:2005}
S.~S.~Adler {\it et al.}  [PHENIX Collaboration],
    Phys.\ Rev.\ D {\bf 71} (2005) 071102
    [arXiv:hep-ex/0502006].

\bibitem{Okada:2005}
K.~Okada  [PHENIX Collaboration], arXiv:hep-ex/0501066.

\bibitem{Abazov:2005wc}
V.M.~Abazov {\it et al.}, [D$\emptyset$ collaboration],
subm. to Phys. Lett. B., Fermilab-Pub-05/523-E,
  [arXiv:hep-ex/0511054].
%The D$\emptyset$ Collaboration, D$\emptyset$note 4859-CONF, %http://www.do.fnal.gov.


\bibitem{EXP}	
M.~McLaughlin {\em et. al.} [E629 collaboration], 
      Phys. Rev. Lett. {\bf 51} (1983) 971;\\
J.~Badier {\em et. al.} [NA3 collaboration], 
      Zeit. f{\"u}r  Physik C {\bf 31} (1986) 341;\\
C.~De Marzo {\em et. al.} [NA24 collaboration], 
      Phys. Rev. D\ {\bf 36} (1987) 8;\\
A.L.S.~Angelis {\em et. al.} [R108 collaboration], 
      Nuc Phys. B {\bf 263} (1986) 228

\bibitem{R806}
E.~Anassontzis {\it et al.} [R806 Collaboration],
      Z.\ Phys.\ C {\bf 13} (1982) 277.

\bibitem{WA70}	
M.~Bonesini {\it et al.}  [WA70 Collaboration],
      Z.\ Phys.\ C {\bf 37} (1988) 535.

\bibitem{R110}
%%%%%%%%%%%%%%%%{Angelis:1989zv}
A.~L.~S.~Angelis {\it et al.}  [CMOR Collaboration],
     Nucl.\ Phys.\ B {\bf 327} (1989) 541.

\bibitem{R807}
%%%%%%%%%%%%%%%%%{Akesson:1989hp}
T.~\AA kesson {\it et al.}  [Axial Field Spectrometer Collaboration],
     Sov.\ J.\ Nucl.\ Phys.\  {\bf 51} (1990) 836
     [Yad.\ Fiz.\  {\bf 51} (1990) 1314].
   
\bibitem{E706}	
 L.~Apanasevich {\em et al.} [E706 collaboration], 	
 Phys. Rev. Lett. {\bf 81} (1998) 2642.

\bibitem{E70604}
%%%%%%%%%%%%%%%%%%%%%{Apanasevich:2004dr}
  L.~Apanasevich {\it et al.}  [E706 Collaboration],
    Phys.\ Rev.\ D {\bf 70} (2004) 092009
    [arXiv:hep-ex/0407011].

\bibitem{UA6}
%%%%%%%%%%%%%%%%%%%%%%{Ballocchi:1998au}
G.~Ballocchi {\it et al.}  [UA6 Collaboration],
     Phys.\ Lett.\ B {\bf 436} (1998) 222.
%%%%%%%%%%%%%%%%%%%%{Werlen:1999ij}
%%%%%%%%%%%%%%%%%M.~Werlen {\it et al.} [UA6 Collaboration],
%%%%%%%%%%%%%%%%%    Phys. \ Lett. \ B {\bf 452} (1999) 201.
    
\bibitem{UA1}
%%%%%%%%%%%%%%%%%%%{Albajar:1988im}
C.~Albajar {\it et al.}  [UA1 Collaboration],
    Phys.\ Lett.\ B {\bf 209} (1988) 385.

\bibitem{UA2}
%%%%%%%%%%%%%%%%%%%{Alitti:1992hn}
J.~Alitti {\it et al.}  [UA2 Collaboration],
    Phys.\ Lett.\ B {\bf 288} (1992) 386.

\bibitem{D02000}
%%%%%%%%%%%%%%%%{Abbott:1999kd}
B.~Abbott {\it et al.}  [D0 Collaboration],
    Phys.\ Rev.\ Lett.\  {\bf 84} (2000) 2786
    [arXiv:hep-ex/9912017].

\bibitem{D02001}
%%%%%%%%%%%%%%%%%%{Abazov:2001af}
V.~M.~Abazov {\it et al.}  [D0 Collaboration],
    Phys.\ Rev.\ Lett.\  {\bf 87} (2001) 251805
    [arXiv:hep-ex/0106026].

\bibitem{CDF2002}
%%%%%%%%%%%%%%%%%%%%%%{Acosta:2002ya}
D.~Acosta {\it et al.}  [CDF Collaboration],
    Phys.\ Rev.\ D {\bf 65} (2002) 112003
    [arXiv:hep-ex/0201004].

\bibitem{CDF2004}
%%%%%%%%%%%%%%%%%%%%%%{Acosta:2004bg} 
D.~Acosta {\it et al.}  [CDF Collaboration],
    Phys.\ Rev.\ D {\bf 70} (2004) 074008
    [arXiv:hep-ex/0404022].
 
\bibitem{cocorico}
P.~Aurenche, T.~Binoth, M.~Fontannaz, J.-P.~Guillet, G.~Heinrich, E.~Pilon 
and M.~Werlen,\\
http://lappweb.in2p3.fr/lapth/PHOX{\_}FAMILY/main.html,\\
http://lappweb.in2p3.fr/lapth/PHOX{\_}FAMILY/jetphox.html

\bibitem{ABDFS}	%\cite{Aurenche:1983ws}
  P.~Aurenche, A.~Douiri, R.~Baier, M.~Fontannaz and D.~Schiff,
  %``Prompt Photon Production At Large P(T) In QCD Beyond The Leading Order,''
  Phys.\ Lett.\ B {\bf 140} (1984) 87;\\
  %%CITATION = PHLTA,B140,87;%%
  P.~Aurenche, R.~Baier, M.~Fontannaz and D.~Schiff,
  %``Prompt Photon Production At Large P(T) Scheme Invariant QCD Predictions And
  %Comparison With Experiment,''
  Nucl.\ Phys.\ B {\bf 297} (1988) 661.

\bibitem{GV}	
L.E.~Gordon and W.~Vogelsang, 
Phys.\ Rev.\ D {\bf 50} (1994) 1901.

\bibitem{ACGG}	
F.~Aversa, P.~Chiappetta, M.~Greco and J.Ph.~Guillet,
Nuc. Phys. B {\bf 327} (1989) 105;\\
P.~Aurenche, P.~Chiappetta, M.~Fontannaz, J.Ph.~Guillet and E.~Pilon, 
Nuc. Phys. B {399} (1993) 34.

%%%%%%%%%%%%\bibitem{ABFOW}		
%%%%%%%%%%%%P.~Aurenche, R.~Baier, M.~Fontannaz, J.F.~Owens and
%%%%%%%%%%%%M.~Werlen, Phys. Rev. D {\bf 39} (1989) 3275.

\bibitem{owens}
H. Baer, J. Ohnemus and J. F. Owens, Phys. \ Lett. B {\bf 234} (1990) 127; 
Phys. \ Rev. \ D {\bf 42} (1990) 61

\bibitem{gordon}
L.~E.~Gordon and W.~Vogelsang,
    Phys.\ Rev.\ D {\bf 52} (1995) 58.

\bibitem{los}
E.~Laenen, G.~Oderda and G.~Sterman,
    Phys.\ Lett.\ B {\bf 438} (1998) 173
    [arXiv:hep-ph/9806467].

\bibitem{cmn} 
S. Catani, M.L. Mangano, P. Nason, 
JHEP {\bf 9807} (1998) 024.

\bibitem{cmnov}
%%%%%%%%%%%%%%%%%%%%%%%%{Catani:1999hs}
S.~Catani, M.L.~Mangano, P.~Nason, C.~Oleari, W.~Vogelsang, 
JHEP {\bf 9903} (1999) 025   
[arXiv:hep-ph/9903436].

\bibitem{ko}
N. Kidonakis, J.F. Owens, 
Phys. \ Rev. \ D {\bf 61} (2000) 094004
[arXiv:hep-ph/9912388].

\bibitem{sv} 
G. Sterman and W. Vogelsang, 
JHEP {\bf 0102} (2001) 016
[arXiv:hep-ph/0011289].

\bibitem{deFlorian:2005wf}
D.~de Florian and W.~Vogelsang,
    Phys.\ Rev.\ D {\bf 72} (2005) 014014
    [arXiv:hep-ph/0506150].

\bibitem{Sterman:2004yk}
G. Sterman and W. Vogelsang, 
Phys. \ Rev. \ D {\bf 71} (2005) 014013
[arXiv:hep-ph/0409234].

\bibitem{Catani:2002ny}
S. Catani, M. Fontannaz, J.-P. Guillet and E. Pilon,
%%%%%%%%%%%%Cross section of isolated prompt photons in hadron hadron collisions,
JHEP {\bf 0205} (2002) 028 [hep-ph/0204023].             

\bibitem{slicing}
M.A. Furman,  
Nucl. Phys. B {\bf 197} (1982) 413;\\ 
W.T. Giele and E.W.N. Glover, 
{Phys. Rev. D} {\bf 46} (1992) 1980;\\
W.T. Giele, E.W.N. Glover and D.A. Kosower, 
{Nucl. Phys. B} {\bf 403} (1993) 633;\\
  P.~Chiappetta, R.~Fergani and J.~P.~Guillet,
  %``Double prompt photon production from hadronic collisions,''
  Phys.\ Lett.\ B {\bf 348} (1995) 646; 
  %%CITATION = PHLTA,B348,646;%%
  %%  P.~Chiappetta, R.~Fergani and J.~P.~Guillet,
  %``Production of two large p(T) hadrons from hadronic collisions,''
  Z.\ Phys.\ C {\bf 69} (1996) 443.
  %%CITATION = ZEPYA,C69,443;%%
  
\bibitem{subtraction}
R.K. Ellis, D.A. Ross and A.E. Terrano, 
{Nucl. Phys. B} {\bf 187} (1981) 421;\\
S. Frixione, Z. Kunszt and A. Signer, 
{Nucl. Phys. B} {\bf 467} (1996) 399;\\
S. Catani and M.H. Seymour, 
{Nucl. Phys. B} {\bf 485} (1997) 291. 

\bibitem{correlations}
Z.~Belghobsi, M.~Fontannaz, J.-P.~Guillet, G.~Heinrich and M.~Werlen, 
in preparation.


\bibitem{cteq6m}
%%%%%%%%%%%%%%%%%%%%%{Pumplin:2002vw}
  J.~Pumplin, D.~R.~Stump, J.~Huston, H.~L.~Lai, P.~Nadolsky and W.~K.~Tung,
  %``New generation of parton distributions with uncertainties from global  QCD
  %analysis,''
  JHEP {\bf 0207} (2002) 012
  [arXiv:hep-ph/0201195].
  %%CITATION = HEP-PH 0201195;%%

\bibitem{BFG}
%%%%%%%%%%%%%%%%\bibitem{Bourhis:1997yu}
  L.~Bourhis, M.~Fontannaz and J.~P.~Guillet,
  %``Quark and gluon fragmentation functions into photons,''
  Eur.\ Phys.\ J.\ C {\bf 2} (1998) 529
  [arXiv:hep-ph/9704447].
  %%CITATION = HEP-PH 9704447;%%       	(1998) 529.
  
\bibitem{martin:2004ir}
A.D.~Martin, R.G.~Roberts,W.J.~Stirling, R.S.~Thorne,
Phys. \ Lett. \ B {\bf 604} (2004) 61
 [arXiv:hep-ph/0410230] 

%\cite{Aktas:2004uv}
\bibitem{h1-2004}
  A.~Aktas {\it et al.}  [H1 Collaboration],
  %``Measurement of prompt photon cross sections in photoproduction at HERA,''
  Eur.\ Phys.\ J.\ C {\bf 38} (2005) 437
  [arXiv:hep-ex/0407018].
  %%CITATION = HEP-EX 0407018;%%

%\cite{Aurenche:1998gv}
\bibitem{afgkpw}
  P.~Aurenche, M.~Fontannaz, J.~P.~Guillet, B.~A.~Kniehl, E.~Pilon and M.~Werlen,
  %``A critical phenomenological study of inclusive photon production in
  %hadronic collisions,''
  Eur.\ Phys.\ J.\ C {\bf 9} (1999) 107
  [arXiv:hep-ph/9811382].
  %%CITATION = HEP-PH 9811382;%%

%\cite{Werlen:1999ij}
\bibitem{Werlen:1999ij}
  M.~Werlen {\it et al.}  [UA6 Collaboration],
  %``A new determination of alpha(s) using direct photon production cross
  %sections in p p and anti-p p collisions at s**(1/2) = 24.3-GeV,''
  Phys.\ Lett.\ B {\bf 452} (1999) 201.

\bibitem{HKKLOT}
J. Huston, E. Kovacs, S. Kuhlmann, H.L. Lai, J.F. Owens and W.K. Tung, 
\PR {D51} (1995) 6139. 
		
\bibitem{Baur:2000xd} 
U.~Baur {\it et. al.} (2000) [arXiv:hep-ph/0005226].	 
		
\bibitem{vv}
%%%%%%%%%%%%%%%%%%%%{Vogelsang:1995bg}
  W.~Vogelsang and A.~Vogt,
  %``Constraints on the proton's gluon distribution from prompt photon
  %production,''
  Nucl.\ Phys.\ B {\bf 453} (1995) 334
  [arXiv:hep-ph/9505404].
  %%CITATION = HEP-PH 9505404;%%

\bibitem{D0jet}
The D$\emptyset$ Collaboration, D$\emptyset$ note 4751-CONF, 
http://www.do.fnal.gov

\bibitem{cdfjet}
A.~Abulencia {\it et.al.} [CDF Collaboration]
[arXiv: hep-ex/0512020]


\end{thebibliography}
\end{document}